\documentclass[aps,prl,twocolumn,showpacs,showkeys,nofootinbib,superscriptaddress]{revtex4-1}

\usepackage[utf8]{inputenc}
\usepackage{graphicx}
\usepackage{amsfonts}
\usepackage{amssymb}
\usepackage{amsmath}
\usepackage[mathscr]{eucal}
\usepackage{setspace}
\usepackage{booktabs}
\usepackage[shellescape,latex]{gmp}
\usempxpackage{amssymb}

\def\sqr#1#2{{\vcenter{\hrule height.#2pt
   \hbox{\vrule width.#2pt height#1pt \kern#1pt
      \vrule width.#2pt}
   \hrule height.#2pt}}}

\def\bsqr#1#2{{\vrule width #1pt height#2pt}}
\def\bsquare{{\mathchoice\bsqr66\bsqr66\bsqr33\bsqr33}}
\def\badbreak{\penalty1000}

\def\fir{{\scriptscriptstyle{\text{\rm IR}}}}                 
\def\fuv{{\scriptscriptstyle{\text{\rm UV}}}}              
\def\cro{{\scriptscriptstyle{\text{\rm A}}}}                 
\def\lm0{{\lambda_0}}                                             
\def\nrN{N}                                                       
\def\cf{\mathfrak{n}}                                         
\def\cfu{\cf_\star}                                              
\def\efN{\mathscr{N}}                                        
\def\efNm{\efN_\star}                                        
\def\w{c}                                                              
\def\v{b}                                                               

\def\edged{\lambda_{1}}    
\def\edgeu{\lambda_{2}}    

\newcommand*{\GtrApprox}{\smallrel\gtrapprox}
\newcommand*{\LessApprox}{\smallrel\lessapprox}

\makeatletter
\newcommand*{\smallrel}[2][.8]{%
  \mathrel{\mathpalette{\smallrel@{#1}}{#2}}%
}
\newcommand*{\smallrel@}[3]{%
  \sbox0{$#2\vcenter{}$}%
  \dimen@=\ht0 %
  \raise\dimen@\hbox{%
    \scalebox{#1}{%
      \raise-\dimen@\hbox{$#2#3\m@th$}%
    }%
  }%
}
\makeatother

\usepackage{amsmath} 
\usepackage{dsfont}  
\usepackage{bm}      

\def\beq{\begin{equation}}
\def\eeq{\end{equation}}
\def\beqs#1\eeqs{\beq\begin{split} #1 \end{split}\eeq}

\long\def\comment#1{}

\usepackage{microtype}
\usepackage[dvipsnames]{xcolor}
\usepackage[colorlinks=true,backref=false, linktocpage=true,
citecolor=PineGreen,urlcolor=PineGreen,linkcolor=PineGreen,pdfpagemode=UseOutlines]{hyperref}

\hypersetup{%
  bookmarksnumbered=true,
  pdftitle = {},
  pdfsubject = {},
  pdfauthor = {},
  pdfkeywords = {}
}

\begin{document}


\title{Unusual Features of QCD Low-Energy Modes in IR Phase}

\author{Andrei\ Alexandru}
\email{aalexan@gwu.edu}
\affiliation{The George Washington University, Washington, DC 20052, USA}

\author{Ivan Horv\'ath}
\email{ihorv2@g.uky.edu}
\affiliation{University of Kentucky, Lexington, KY 40506, USA}

\date{Jun 2, 2021}

\begin{abstract}

\comment{
We recently proposed that there is a phase in thermal QCD ({\em IR phase})
at temperatures $T>T_\fir$ well above the chiral crossover, featuring the
elements of scale invariance in the infrared (IR). Here we evaluate spatial 
dimensions of low-energy Dirac modes in this phase. 
The needed concept of effective dimension $d_\fir$ reflective of IR features has 
emerged recently, and is well-founded. We find low-$d_\fir$ ($< \!3$) modes in 
spectral range $0 \le \edged < \lambda < \edgeu$, concluding that 
$\edged \!=\! \lambda_\fir \!=\! 0$, where $\lambda_\fir$ is the defining singularity of 
IR phase, and that $\edgeu \!=\! \lambda_A$, where $\lambda_A \!>\! T$ is the previously 
seen Anderson-like point. 
Details near $\lambda_\fir$ are unexpected in that only exact zero modes have 
$d_\fir \!=\! 3$, while a thin spectral layer near zero has $d_\fir \!=\! 2$, followed by an 
extended layer of $d_\fir \!=\! 1$ modes. With only integers appearing, $d_\fir$ may have 
topological meaning near $\lambda_\fir$. 
We find similar structure at $\lambda_A$, and associate its adjacent thin layer 
($d_\fir \!\GtrApprox 2$) with Anderson-like criticality. 
Our analysis shows that IR phase has signature non-analyticities of $d_\fir(\lambda)$ 
at $\lambda_\fir$ and $\lambda_A$. The role of $\lambda_A$ in previously conjectured 
decoupling of IR component is explained. The obtained
dimension structure may be important for faithful model descriptions of near-perfect
fluidity observed in accelerator experiments.
}


\comment{It was recently proposed that there is a phase in thermal QCD ({\em IR phase}) 
at temperatures well above the chiral crossover, featuring elements of scale 
invariance in the infrared (IR). Here we study IR scaling of effective volumes 
occupied by Dirac modes in this phase. The quantity describing it, $d_\fir$, 
has meaning analogous to measure-based definitions of dimension 
(Hausdorff, Minkowski). Ordinary extended modes, such as those at high energy, 
have $d_\fir \!=\! 3$. We find $d_\fir \!<\! 3$ in the spectral range whose 
lower edge coincides with $\lambda_\fir \!=\!0$, the singularity of spectral 
density defining the IR phase, and the upper edge with $\lambda_A$, the previously 
identified Anderson-like non-analyticity. 
Details near $\lambda_\fir$ are unexpected in that only exact zero modes are 
$d_\fir \!=\! 3$, while a thin spectral layer near zero is $d_\fir \!=\! 2$, 
followed by an extended layer of $d_\fir \!=\! 1$ modes. With only integer
values appearing, $d_\fir$ may have topological origin. 
We find similar structure at $\lambda_A$, and associate its adjacent thin layer 
($d_\fir \!\GtrApprox 2$) with Anderson-like criticality. 
Our analysis reveals the manner in which non-analyticities at $\lambda_\fir$ 
and $\lambda_A$, defined by other means, appear in $d_\fir(\lambda)$.
This dimension structure may be important for understanding the near-perfect 
fluidity of the quark-gluon medium seen in accelerator experiments. 
The role of $\lambda_A$ in previously conjectured decoupling of IR 
component is explained.}

It was recently proposed that there is a phase in thermal QCD ({\em IR phase}) 
at temperatures well above the chiral crossover, featuring elements of scale 
invariance in the infrared (IR). 
Here we study the effective spatial dimensions, $d_\fir$, of Dirac low-energy modes in 
this phase, in the context of pure-glue QCD. 
Our $d_\fir$ is based on the scaling of mode support toward thermodynamic 
limit, and hence is an IR probe. 
Ordinary extended modes, such as those at high energy, 
have $d_\fir \!=\! 3$. We find $d_\fir \!<\! 3$ in the spectral range whose 
lower edge coincides with $\lambda_\fir \!=\!0$, the singularity of spectral 
density defining the IR phase, and the upper edge with $\lambda_A$, the previously 
identified Anderson-like non-analyticity. 
Details near $\lambda_\fir$ are unexpected in that only exact zero modes are 
$d_\fir \!=\! 3$, while a thin spectral layer near zero is $d_\fir \!=\! 2$, 
followed by an extended layer of $d_\fir \!=\! 1$ modes. With only integer
values appearing, $d_\fir$ may have topological origin. 
We find similar structure at $\lambda_A$, and associate its adjacent thin layer 
($d_\fir \!\GtrApprox 2$) with Anderson-like criticality. 
Our analysis reveals the manner in which non-analyticities at $\lambda_\fir$ 
and $\lambda_A$, originally identified in other quantities, appear in $d_\fir(\lambda)$.
This dimension structure may be important for understanding the near-perfect 
fluidity of the quark-gluon medium seen in accelerator experiments. 
The role of $\lambda_A$ in previously conjectured decoupling of IR 
component is explained.

\medskip

\end{abstract}

\maketitle


\noindent
{\bf 1.~Introduction. $\;$}The interest in Dirac eigenmodes~of Euclidean Quantum 
Chromodynamics (QCD) has a long history, sparked in part by the role of zero modes 
in topology of gauge fields ($\eta^\prime$ problem~\cite{tHooft:1976rip}) and by that 
of near-zero modes in spontaneous chiral symmetry breaking (Banks-Casher 
relation~\cite{Banks:1979yr}). While modeling of low-energy QCD based on 
instantons could qualitatively accommodate these features~\cite{Schafer:1996wv}, 
the birth of numerical lattice QCD~\cite{Creutz:1979dw} allowed for computation of 
Dirac eigenmodes from first principles~\cite{Smit:1986fn, Smit:1987jh}. This provided 
access to details of their true structure and thus an important window into the inner 
workings of QCD dynamics (see 
e.g.~\cite{Edwards:1999zm, Horvath:2002gk, Alexandru:2012sd, Giordano:2013taa, 
Alexandru:2014paa, Alexandru:2015fxa}).

Recently, Dirac eigenmodes were used to infer the existence of a new phase in 
thermal QCD (IR phase)~\cite{Alexandru:2019gdm}, showing certain signs of scale 
invariance at energies below temperature $T$. It was proposed that, past a crossover 
region (chiral $T_c \!\approx\!$ 155 MeV), a true  QCD phase transition may occur 
at temperature $T_\fir$ (200-250 MeV), marking the restoration of scale invariance 
in the infrared (IR). The reasoning was based on the proposition that the observed 
power law behavior of Dirac spectral density in IR 
($\rho(\lambda) \approx 1/\lambda$) arises 
due to the underlying IR scale invariance of glue fields. This is corroborated by 
the finding that such Dirac spectral feature also occurs near SU(3) conformal window 
at zero temperature~\cite{Alexandru:2014zna,Alexandru:2014paa}, placing both 
corners of the theory parameter space into one contiguous dynamical regime, 
the IR phase. 

Given the relevance of the above to the physics of quark-gluon plasma studied at RHIC
and LHC, as well as to the physics of the early universe (see e.g. \cite{Busza:2018rrf} 
for review), our aim in this work is to describe IR phase in a manner that sheds more 
light on the mechanism of its conjectured scale invariance. Owing to their proposed common 
origin, the new insight would also be valuable for understanding the mechanism of 
scale invariance in the strongly coupled part of conformal window.

To work with a concrete problem, consider pure glue QCD at $T\!>\!T_\fir$. 
Ref.~\cite{Alexandru:2019gdm} proposes the existence of a physical energy 
scale $\Lambda_\fir \!=\! \Lambda_\fir(T) \lessapprox T$, such that the theory is scale 
invariant at energies $E \!<\! \Lambda_\fir$. In this scenario, gauge coupling stops 
running below $\Lambda_\fir$, leading to non-analyticity at this point. How could
such feature arise in QCD? Motivated by clean bimodality of $\rho(\lambda)$, 
Ref.~\cite{Alexandru:2019gdm} suggested that IR gauge fields decouple 
and fluctuate independently of the bulk in the IR phase. The ensuing mismatch 
between the two independent components of the system can 
then produce such non-analyticity. 

In what follows we make this proposal more precise and concrete by focusing 
on non-analyticities of Dirac spectra. After all, a true non-analyticity of running
coupling would be reflected in all dynamical elements of the theory, especially in 
a Dirac eigensystem where the singularity of $\rho(\lambda)$ at 
$\lambda_\fir \!=\!0$ 
first suggested the existence of IR phase. Hence, we aim to identify these 
non-analyticities and to determine their role in the above dynamical scenario.
   
In addition to $\lambda_\fir \!=\!0$ ($\rho \! \approx \! 1/\lambda$), there is 
a well-known singularity of $\rho(\lambda)$ at 
$\lambda_\fuv \!=\! \infty$ ($\rho \!\approx\! \lambda^3$).  Moreover, a different 
type of spectral non-analyticity, namely the Anderson-like localization point 
$\lambda_\cro \!>\!T$, has been advocated for and studied
for some time~\cite{GarciaGarcia:2006gr, Kovacs:2010wx, Giordano:2013taa}. 
While the singularity at $\lambda_\fuv$ is expected at all temperatures 
(asymptotic freedom), $\lambda_A$ appears to be a companion of IR phase. 
If so, then $\lambda_\fir$ and $\lambda_A$ should be associated in some 
manner. 

To identify such possible connection, as well as to search for additional 
non-analyticities of Dirac spectra, we study the dimension $d_\fir\!=\!d_\fir(\lambda)$ 
of spatial region effectively occupied by modes. We emphasize that we use 
the infrared concept of dimension which probes the response of effective 
volume to the release of IR cutoff. This approach has its roots in
Ref.~\cite{Horvath:2018aap} and will be discussed in Sec.~2.\footnote{The full 
account of dimension theory will be given in Ref.~\cite{Hor2021_B}.}
In Sec.~3 we describe our numerical results. Their proposed implications are 
elaborated upon~in~Sec.~4. 

This work is concerned with results in pure-glue QCD. However, we believe that 
our findings carry over in the presence of dynamical quarks. This is only 
a conjecture at this stage, albeit a reasonable one given the likely 
topological nature of the reported features. We plan to investigate this issue 
in future studies.


\begin{figure}[t]
\begin{center}
    \vskip -0.05in
    \centerline{
    \hskip 0.1in
    \includegraphics[width=9.0truecm,angle=0]{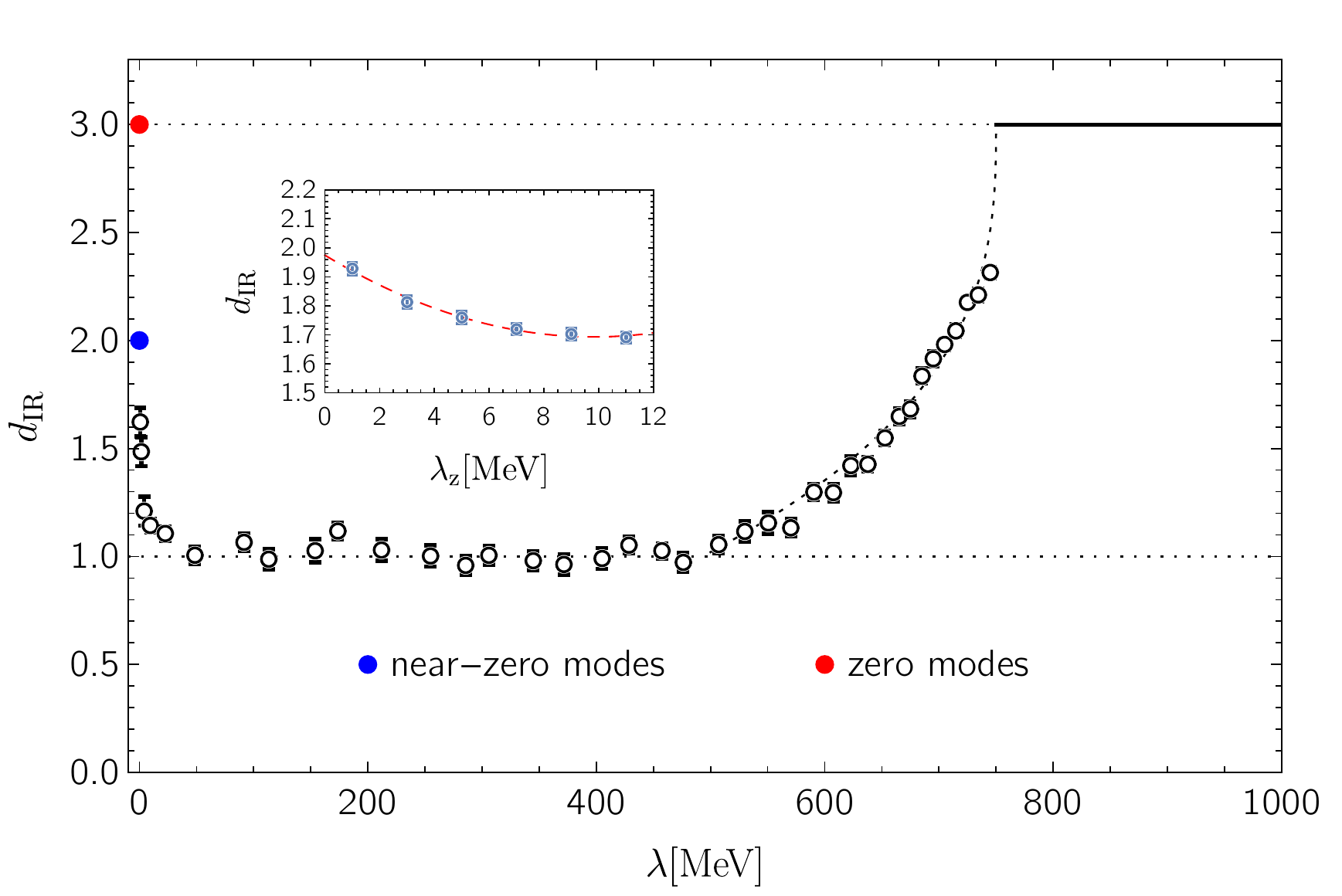}
    }
    \vskip -0.05in
    \caption{IR dimension of Dirac modes at $T=1.12\,T_\fir$.
     }
    \label{fig:dims}
    \vskip -0.45in
\end{center}
\end{figure}

\smallskip

\noindent
{\bf 2. Dimensions.} 
Consider thermal QCD~in 4-volume $L^3/T$, regularized on a 
hypercubic lattice with spacing~$a$. For questions of thermodynamic 
($L \!\to\! \infty$) limit, scale $1/L$ is the IR cutoff, similarly to $1/a$ being 
the UV cutoff. The eigenmodes $D \psi_\lambda(x) \!=\! i \lambda \psi_\lambda(x)$ 
of a continuum Dirac operator $D$ have suitable lattice counterparts, supplied here 
by those of the overlap operator~\cite{Neuberger:1997fp}. This ensures continuum-like 
chiral and topological properties.
  
Following~\cite{Horvath:2018aap, Hor2021_B}, we assign spatial dimensions 
to Dirac eigenmodes as follows. Consider the eigenmode 
$\psi_\lambda \!=\! \psi_\lambda(x_i, L, a)$ at given $\lambda$ and cutoffs. 
Although nominally such mode extends over all $N \!=\! (L/a)^3/(Ta)$ sites, this
``counting" should be modified if the probabilities
$P \!=\! (p_1,p_2,\ldots,p_N)$, $p_i \!=\! \psi_\lambda^+ \psi_\lambda(x_i)$, are 
effectively concentrated in fewer ($\efN < N$) sites. Assume that such effective 
counting $\efN \!=\! \efN[P]$ is in place. While $N \propto L^3$ at fixed $a$, 
the effective $\efN$ is governed by $\efN \propto L^{d_\fir}$ for $L \to \infty$, with 
$0 \le d_\fir \le 3$. Dimension $d_\fir \!=\! d_\fir(a)$ is ``infrared" since it probes 
the removal of IR cutoff. 
Similarly, while $N \!\propto\! a^{-4}$, the power $d_\fuv \!=\! d_\fuv(L)$ 
in $\efN \propto a^{-d_\fuv}$ for $a \!\to\! 0$ ($0 \le d_\fuv \le 4$) is
the effective UV dimension.
 
Effective number theory~\cite{Horvath:2018aap} specifies all additive effective 
counting schemes $\efN$, and leads to 
\begin{equation}
      \efNm[P]  \,=\, \sum_{i=1}^\nrN \cfu(\nrN p_i)   \quad,\quad
      \cfu(\w)  \; = \;   \min\, \{ \w, 1 \}    \;
      \label{eq:022}         
\end{equation}
for a scheme that consistently delimits the effective subset (support) of objects 
with probabilities~\cite{Hor2021_B}. This makes $d_\fir$, $d_\fuv$ well-defined 
characteristics of mode's effective support. Since $\efNm$ is additive, $d_\fuv$ 
and $d_\fir$ are stochastic measure-based constructs analogous to box-counting and 
Hausdorff dimensions~\cite{Hor2021_B}.\footnote{Ref.~\cite{Aubin:2004mp}
describes a similar approach based on the participation number. Due to the lack 
of additivity, it is not Hausdorff-like.}
In fact, the method can be adapted to define dimensions of fixed (non-stochastic) 
fractal sets, and we verified in few cases that it produces $d_\fuv$ consistent with 
their Hausdorff dimensions.

Here we focus on IR dimensions of Dirac modes. Since $\psi_\lambda$ is 
a statistical object, $d_\fir$ is defined by
\begin{equation}
     \langle \, \efNm \,\rangle_{a,L,\lambda} \,\propto\, L^{d_\fir(a,\lambda)} 
     \quad \text{for} \quad L \to \infty  \quad 
     \label{eq:027}	
\end{equation} 
where $\langle \cdots \rangle_{a,L,\lambda}$ denotes the QCD average at fixed 
cutoffs, and the spectral average in the vicinity of $\lambda$. 

\smallskip

\noindent
{\bf 3. The Results.} 
We computed $d_\fir(\lambda)$ in IR phase of pure glue QCD ($T\!=\!1.12\, T_\fir$) 
using the setup of Ref.~\cite{Alexandru:2019gdm}
(Wilson action at $\beta \!=\! 6.054$, $a\!=\!0.085\,$fm via $r_0\!=\!0.5\,$fm, 
overlap operator at $\rho \!=\! 26/19$). Systems with 
$L/a \!=\! 16,20,24,32,40,48,64$ and $1/(Ta)\!=\!7$ were analyzed.
Special care was required to reliably identify the zero modes, 
and to implement the overap operator in a numerically 
efficient way~\cite{Alexandru:2011sc, Alexandru:2014mqy, Alexandru:2011ee}.

Our main results are conveyed by Fig.~\ref{fig:dims}. Dimensions 
were obtained from fits to the asymptotic form~\eqref{eq:027} using four largest 
systems (range 2.7--5.4$\,$fm) in the analysis. The fits had statistically acceptable 
$\chi^2$/dof in general. The spectral intervals associated with plotted points are 
disjoint, and cover the region shown. 

We find the spectral interval $( \edged, \edgeu )$ of low-$d_\fir$ modes
($d_\fir \!<\! 3$), featuring three regimes:  the constant central plateau at 
$d_\fir \!\approx\!1$ and two half-peaks on the sides (``left rise" and ``right rise"). 
The details are as follows.

\begin{figure}[t]
\begin{center}
    \vskip -0.05in
    \centerline{
    \hskip 0.1in
    \includegraphics[width=9.0truecm,angle=0]{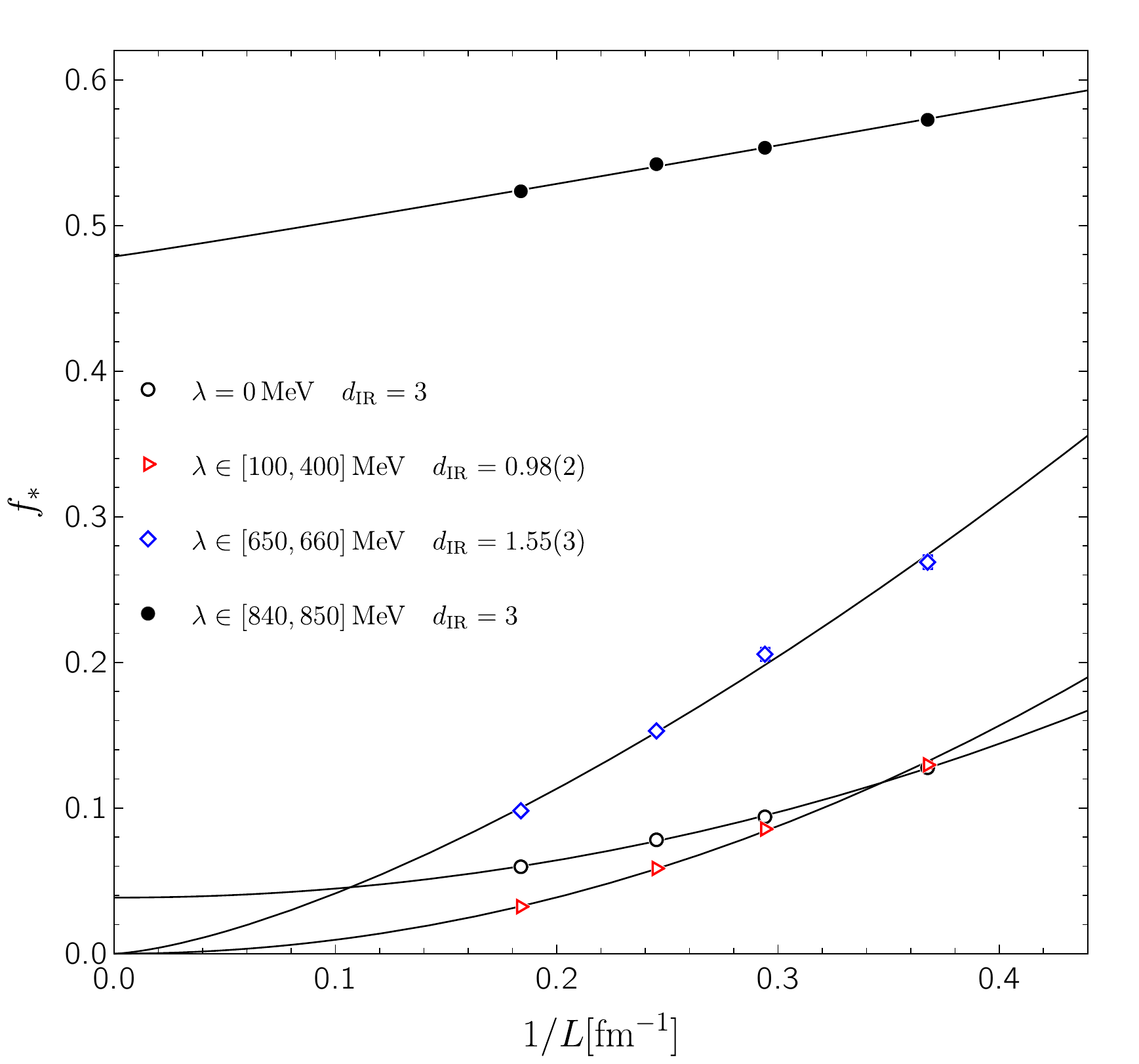}
    }
    \vskip -0.05in
    \caption{The fraction $f_\star$ of space occupied by modes vs $1/L$.
     }
    \label{fig:pfits}
    \vskip -0.40in
\end{center}
\end{figure}

\noindent
{\bf Zero Modes.} The overlap operator supports exact topological zero modes, 
separated from the rest of the spectrum in any finite volume. We find that
$d_\fir(0)\!=\!3$ (red point in Fig.~\ref{fig:dims}), with the behavior of 
$f_\star \!=\! \efNm/\nrN \propto L^{d_\fir-3}$ shown in Fig.~\ref{fig:pfits}. 
Using the Anderson localization terminology, this means that zeromodes are 
``extended" since they occupy a finite fraction of volume ($\approx\,$4\%) in 
$L \!\to\! \infty$ limit.

\noindent
{\bf Near-Zero Modes.} 
Is there a layer of deep IR modes akin to strict zeromodes in terms of $d_\fir$? 
To assess this, we treat the extent $\lambda_{\text{z}}$ of such potential 
region $(0,\lambda_{\text{z}})$ as a parameter, and evaluate the average $d_\fir$ 
of modes from this interval. The result is shown in the inset of Fig.~\ref{fig:dims}. 
It reveals that, rather than $d_\fir\!=\!3$, the deepest IR modes afforded by our 
data approach $d_\fir\!=\!2$, and hence $\edged \!=\!0$. In the absence of 
evidence for its finite width, we 
represent the $d_\fir\!=\!2$ layer by a single (blue) point adjacent to point
representing zeromodes.  Note that, although $d_\fir \!<\!3$ implies that 
quarks in these modes occupy a space of measure zero relative to the entire 
volume, they are not necessarily localized in terms of
distances involved.

\begin{figure}[b]
\begin{center}
    \vskip -0.35in
    \centerline{
    \hskip 0.1cm
    \includegraphics[width=9truecm,angle=0]{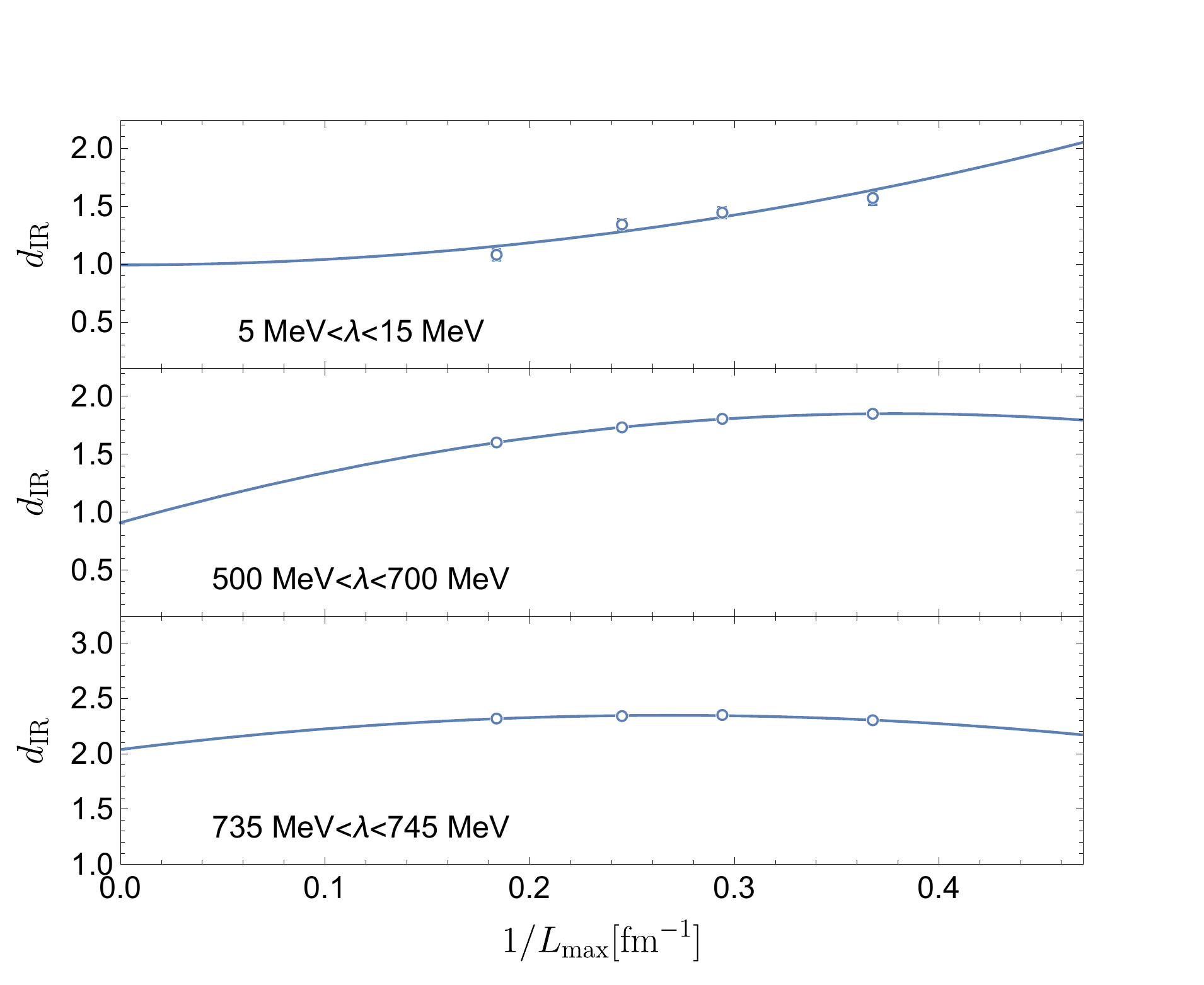}
    }
    \vskip -0.05in
    \caption{$d_\fir(L_{\text{max}})$ as a function $1/L_{\text{max}}$ in left and right rises.
     }
    \label{fig:dimtrends}
    \vskip -0.40in
\end{center}
\end{figure}

\noindent
{\bf The Left Rise.} Consider now the entire left rise in~Fig.~\ref{fig:dims}. 
Since its spectral extent ends just above $1/L_{\text{max}}$ 
($L_{\text{max}} \!=\! 64a$), we need to ask whether the onset of the rise 
marks a non-zero IR scale or vanishes with IR cutoff. To assess this, we vary 
$L_{\text{max}}$ by using sizes of smaller simulated systems as IR cutoffs in 
the dimension analysis. Fig.~\ref{fig:dimtrends} (left) shows the dependence
of $d_\fir$ on $L_{\text{max}}$ for spectral range $5\!-\!15\,$MeV inside 
the rise. We observe lowering of 
the dimension toward $d\fir \!\approx\! 1$ with increasing $L_{\text{max}}$. 
The consistency of such trends makes us to suggest that the width of the left 
rise vanishes in thermodynamic limit and a new dynamical scale is not generated.

\begin{figure}[b]
\begin{center}
    \vskip -0.35in
    \centerline{
    \hskip 0.1in
    \includegraphics[width=9truecm,angle=0]{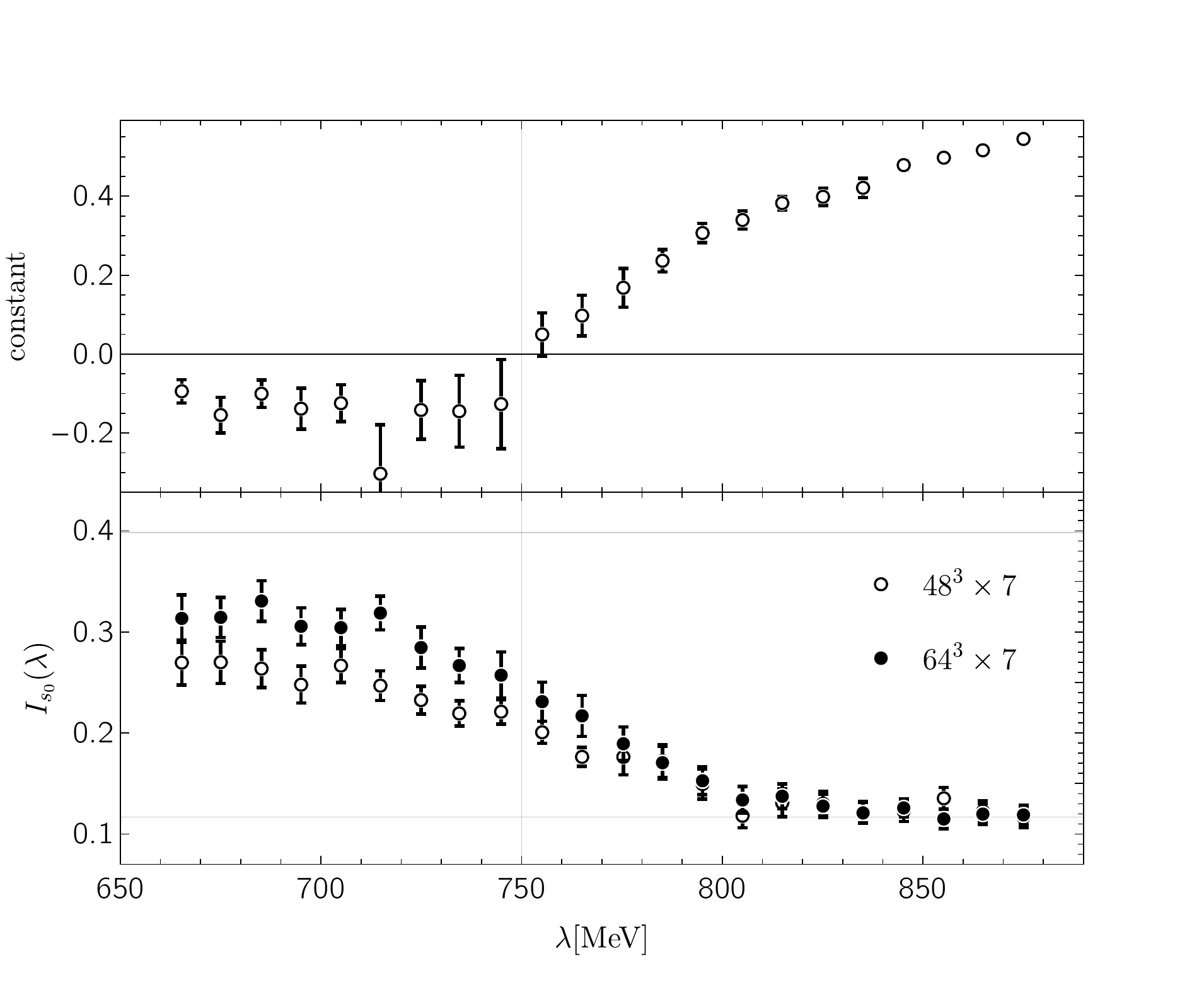}
    }
    \vskip -0.10in
    \caption{The value of constant term in fits of $f_\star(1/L)$ (top) and the value
    of $I_{0.508}$ (bottom) in the vicinity of $\lambda_2$.
     }
    \label{fig:upedge}
    \vskip -0.40in
\end{center}
\end{figure}

\noindent
{\bf The Upper Edge.} The low-$d_\fir$ regime in Fig.~\ref{fig:dims} ends 
abruptly at $\edgeu \!\approx \! 750\,$MeV, suggesting a non-analytic behavior. 
The position of the edge can be estimated by including a constant 
in fits of $f_\star$ (0-th power in addition to floating positive power). 
Fig.~\ref{fig:upedge} (top) shows this constant in the relevant spectral range. 
The small negative values below $\approx \!750\,$MeV indicate that the term 
is redundant since the leading contribution has to be positive on geometric 
grounds. Hence, $d_\fir \!< 3$ in this region.  A statistically significant positive 
value, on the other hand, implies $d_\fir \!=\! 3$ and the extended modes. The 
approach to constant $f_\star > 0$ in the extended regime (black horizontal 
line in Fig.~\ref{fig:dims}) is exemplified in Fig.~\ref{fig:pfits} (full circles). 

The existence of Anderson-like point in QCD Dirac spectra, defined as $\lambda_A$ 
where the statistics of unfolded level spacings changes from Poisson  
($\lambda \!<\! \lambda_A$) to Dyson-Wigner ($\lambda \!>\! \lambda_A$), 
has been concluded in Ref.~\cite{Giordano:2013taa}. To check whether 
the low-$d_\fir$ upper edge $\edgeu$ can be identified with $\lambda_A$, 
we use their procedure and look for the transition in 
\begin{equation}
        I_{s_0}(\lambda) = \int_0^{s_0} ds \,p(s, \lambda) 
\end{equation}
with $p(s,\lambda)$ the distribution of level spacings in the vicinity of $\lambda$.
The value $s_0 \!\approx\! 0.508$ maximizes the difference in $I_{s_0}$ for 
Poisson ($I_{0.508} \!\approx \! 0.398$) and Dyson-Wigner unitary class 
($I_{0.508} \!\approx \! 0.117$) statistics. 
Fig.~\ref{fig:upedge} (bottom) shows $I_{0.508}(\lambda)$ for our two largest 
lattices. The approach to constant Dyson-Wigner value on the right is 
clear and the volume trend on the left is in the expected direction toward 
the Poisson value.
The transition region is rather wide but approximately centered around $750\,$MeV.
These results suggest that $\edgeu$ and $\lambda_A$ represent the same 
spectral point in thermodynamic limit. 

\noindent
{\bf The Right Rise.} The example of $f_\star(1/L)$ inside the right rise is shown in 
Fig.~\ref{fig:pfits}. Like for the left rise, we need to inquire whether the spectral 
width of this feature remains finite in thermodynamic limit. Fig.~\ref{fig:dimtrends} 
(middle) shows $d_\fir$ versus floating $1/L_{\text{max}}$ for the spectral range 
$500\!-\!700\,$MeV extending almost across the rise. A clear decrease of $d_\fir$ with 
increasing $L_{\text{max}}$ is seen. The observed change is well described by 
quadratic polynomial, leading to $L_{\text{max}} \!\to\! \infty$ value consistent 
with that of the plateau ($d_\fir \!\approx\!1$). On the other hand, very close to 
the edge $(735 \!-\! 745)\,$MeV, the same fitting suggests $d_\fir \!\GtrApprox\! 2$. 
This leads us to propose that the left and right rises are similar in that the spectral 
widths of these features vanish in thermodynamic limit, each morphing into 
a point-like layer of well-defined lower dimension. 

\smallskip

\noindent
{\bf 4. The Synthesis and Discussion.} 
The analytic structure of Dirac eigensystem can be useful for detecting the phases 
of QCD. For example, entering the IR phase by crossing $T_\fir$ is characterized by 
adding the IR power singularity of mode density $\rho(\lambda)$ to its already present 
UV singularity~\cite{Alexandru:2019gdm}. 
The $\lambda^3$ divergence at $\lambda_\fuv \!=\! \infty$ reflects the strict power 
law $\rho(\lambda) \!=\! c \lambda^3$ directly at the Gaussian UV fixed point. 
Similarly, we associate the $1/\lambda$ singularity at $\lambda_\fir \!=\!0$ 
with the power law at the strongly-coupled IR fixed point governing the IR 
component of the system~\cite{Alexandru:2019gdm}. 

\begin{figure}[t]
\begin{center}
    \centerline{
    \hskip 0.1in
    \includegraphics[width=8.1truecm,angle=0]{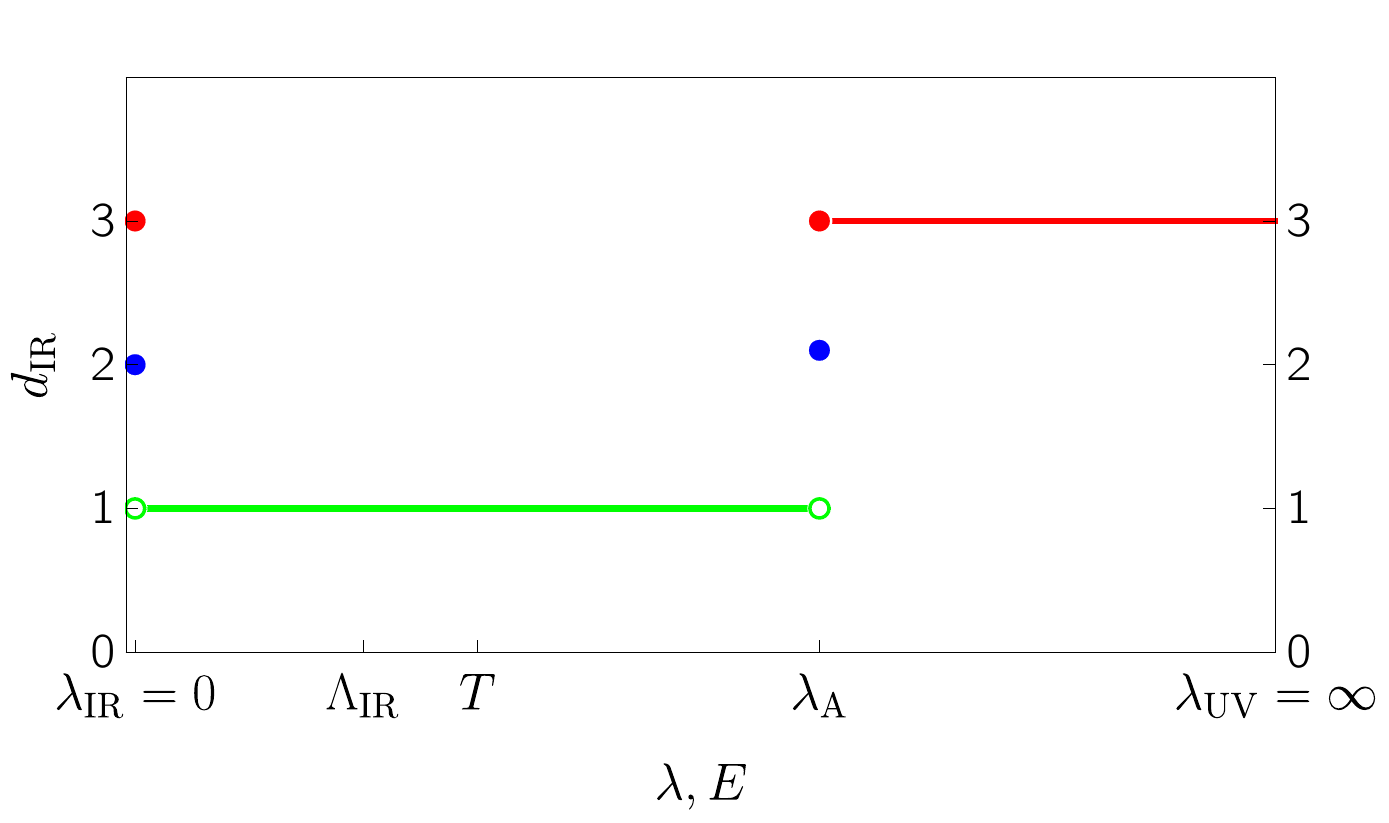}
    }
    \vskip -0.05in
    \caption{Dimension $d_\fir$ of Dirac modes in the IR phase. The abscissa is
    common to physical energy scale $E$ and the Dirac~$\lambda$. }
    \label{fig:dims_illus}
    \vspace{-0.40in}
\end{center}
\end{figure}

Here we extended this analyticity angle by studying the spatial dimension 
$d_\fir(\lambda)$ of Dirac modes. The results lead us to propose that 
the IR phase of QCD is characterized by the existence of a spectral range 
$0 \!\le\! \edged \!<\! \lambda \!<\! \edgeu$ of low-$d_\fir$ modes. 
The edges $\edged$, $\edgeu$ are non-analyticities of $d_\fir(\lambda)$. Our data 
is consistent with the following infinite-volume predictions. 
{\bf (P1)} $d_\fir \!=\! 3$ for zero modes. 
{\bf (P2)} There is a point-like layer of $d_\fir \!=\! 2$ near-zero modes and hence
$\edged \!=\! \lambda_\fir \!=\! 0$. We are not aware of any model that predicts this.
{\bf (P3)} A plateau of $d_\fir \!\approx\! 1$ modes spans the interior of 
low-$d_\fir$ interval.
{\bf (P4)} $\edgeu \!=\! \lambda_A$, where $\lambda_A$ is the Anderson-like 
transition point. 
{\bf (P5)} A point-like layer of $d_\fir \!\GtrApprox\!2$ modes exists on 
the inner side of $\edgeu$. The resulting $d_\fir(\lambda)$ is shown in 
Fig.~\ref{fig:dims_illus}. 

The study of $\rho(\lambda)$ and $d_\fir(\lambda)$ thus offers at least three 
Dirac non-analyticities $\lambda_\fir \!=\! \edged$, 
$\lambda_A \!=\! \edgeu$, $\lambda_\fuv$ in IR phase, with scales 
$T$ and $\Lambda_\fir(T)$ inside the low-$d_\fir$ range
\begin{equation}
    0 = \lambda_\fir < \Lambda_\fir < T < \lambda_A < \lambda_\fuv = \infty
\end{equation}
While the presence of $\lambda_\fir$ as a singularity in $\rho(\lambda)$ 
is natural (see above), the purpose of $\lambda_A$ may seem puzzling. 
What is the role of a ``phase transition" in the internal parameter 
$\lambda$ of the theory? Treating a QCD-induced Dirac 
dynamics near $\lambda_A$ as a model description of some system, the
phase transition at $\lambda_A$ means that its ``states" at 
$\lambda \!<\! \lambda_A$ are unrelated to those at $\lambda \!>\! \lambda_A$, 
which is further underlined here by the former being low-$d_\fir$. 
We ascribe this mutual independence, realized by non-analyticity at 
$\lambda_A$, to the decoupling of IR glue fields from the bulk.
Hence, $\lambda_A$ is connected to the proposed mechanism of strongly-coupled 
IR scale invariance~\cite{Alexandru:2019gdm}, naturally tying with 
the non-analyticity of running coupling at $\Lambda_\fir$.

We wish to make the following additional remarks.
{\bf (a)} The structure of $d_\fir(\lambda)$ is the same at $\lambda_\fir$ and 
$\lambda_A$, involving a two-step non-analytic transition from nominal 
$d_\fir \!=\!3$ to low-$d_\fir$ regime.
{\bf (b)} The existence of the ``middle point" at $\lambda_A$ is natural upon 
accepting its Anderson nature. Indeed, we associate it with the critical region
in Anderson models~\cite{Evers_2008}, which shrink to zero width in thermodynamic 
limit. 
The observed $2 \LessApprox d_\fir(\lambda_A) \!<\! 3$ describes the system 
at strict criticality. 
{\bf (c)} However, $d_\fir \!\approx\! 1$ below $\lambda_A$ is at odds with 
exponential localization in Anderson models. Clarifying this discrepancy is 
an interesting issue to resolve.
{\bf (d)} Despite $d_\fir$ being Hausdorff-like, only near-integer dimensions 
appear around $\lambda_\fir$. Hence, IR dimension may have topological 
meaning for IR modes. The structure near $\lambda_\fir$ is interesting in that 
{\em all} ``topological dimensions" occur.
{\bf (e)} The appearance of $d_\fir \!=\! 2 , 1$ near $\lambda_\fir$ 
is unexpected in light of simple instanton-based models~\cite{Schafer:1996wv}. 
Adding this feature to model descriptions of IR component could be valuable 
since the latter may capture the near-perfect fluidity properties of 
the strongly-interacting medium seen at accelerator 
experiments~\cite{Alexandru:2019gdm}.
{\bf (f)} The proposal for decoupling of IR component used the UV-IR bimodality 
of Dirac spectral density as its corroborating feature~\cite{Alexandru:2019gdm}.  
Since the position 
$0 \!<\! \lambda_{\text{min}} \!< T$ of minimal $\rho(\lambda)$ plays a special
role here, it is desirable to clarify its status regarding analyticity.
{\bf (g)} Albeit indirectly and from different angles, recent works involving 
thermal Dirac modes~\cite{Ding:2020xlj,Aoki:2020noz, Vig:2021oyt} provide 
an additional information on IR phase.
{\bf (h)} We are not aware of a theoretical argument or lattice evidence at this
time, suggesting that $d_\fuv(\lambda,T) < 4$ for any $\lambda,\,T$ in QCD.

\begin{acknowledgments}
A.A. is supported in part by the U.S. DOE Grant No. DE-FG02-95ER40907.
I.H. acknowledges the discussions with Peter Marko\v{s} and Robert Mendris. 
\end{acknowledgments}

\bibliography{my-references}

\end{document}